# Scale properties as a basis of power law relaxation processes

A. Fondado, J. Mira, and J. Rivas
Departamento de Física Aplicada, Universidade de Santiago de Compostela
E-15782 Santiago de Compostela, Spain

Computer simulations of first-order relaxation processes show that the spatial configurations of the system acquire an invariant shape once the stationary regime is attained. Inspired by them we find that, in any first-order relaxation process, if the interaction that governs the system fulfils a simple scale property, then the relaxation will end up by following a stationary process described by a power law. A scaling law and some invariants are obtained for the time evolution of the system in such a case.

76.20.+q, 76.90.+d, 75.40.Mg

There are plenty of examples of systems with time-varying properties both in nature and in the economical and social world.[1] Among them relaxation processes with long-term-decay behavior occupy a privileged place. The natural tendency of such systems to an equilibrium state makes them quite often to approach a stationary regime, where the parameters describing their evolution are time-independent.

Much work in this field has been done on magnetic materials as they provide a paradigmatic scenario[2] easy to manipulate from the experimental point of view, like for example in spin-glasses,[3] which present well defined dynamics. Assemblies of magnetic interacting particles have also been studied both experimentally[4] and by Monte Carlo simulations,[5] some of which showed that the relaxation rate of such assemblies follows a universal power law.[6] This prediction has been confirmed by measurements of relaxation in granular magnetic films.[7,8]

These power law dynamics seem to be a point in common with many relaxation processes, going from magnetic materials or earthquakes[9] to the spatial redistribution of individuals.[10] It has also been demonstrated that they describe the stationary regime of the free relaxation of systems described by the general diffusion-like equation[11]

$$\Theta \mathbf{X} - \frac{\partial \mathbf{X}}{\partial t} = 0 \qquad (1)$$

where $\mathbf{X}$ is the field that describes microscopically the state of a system that is changing under the effect of an interaction described by any spatial-like operator $\Theta$ (i.e.: the time derivative of first order of $\mathbf{X}$ is only a function of the non-time dependence of $\mathbf{X}$). If we let the scalar magnitude $\psi$ be any kind of average calculated over the values of $\mathbf{X}$

(usually $\psi$ is a measurable magnitude used to describe the system macroscopically) and we assume that $\psi$ is monotonically decreasing (positive) and that $\lim_{t\to\infty}\psi(t)=0$, it was demonstrated[11] that a process is stationary if there exists a function $f$ fulfilling

$$\frac{\psi'_1}{\psi'_2} = f\left(\frac{\psi_1}{\psi_2}\right) \qquad (2)$$

(with $\psi'$ the time derivative of $\psi$) and also that in such cases the relaxation is described by a power law

$$\psi = \psi_0\left(1+\frac{t}{\gamma\tau}\right)^{-\gamma} \qquad (3)$$

equation that yields the well-known Debye equation for the particular limit $\gamma\to\infty$

$$\psi_D = \psi_0 e^{-\frac{t}{\tau}}. \qquad (4)$$

As $\Theta$ represents the interaction it is essential to investigate which properties of this operator lead to a stationary regime. It must be kept in mind that the step from Eq. (1) to Eq. (3) is far from straightforward, specially if we take into account non-linear interactions.

For this purpose we have paid attention to the microscopic details of the stationary state by examining configurations taken along time from computer models, in order to obtain the stationary regime in a reasonable time. They consist of a lineal chain of $N$ elements $x_i$, that relax through non-linear interaction with nearest neighbours. The elements $x_0$ and $x_{n-1}$, interact among them, therefore the system can be interpreted as a periodic chain of period $N$ or a ring. In each iteration (time increase $\Delta t$) the increment of $x_i$ is

$$\Delta x_i = \Delta t\left[(x_{i+1}-x_i)|x_{i+1}-x_i|^{\lambda-1} + (x_{i-1}-x_i)|x_{i-1}-x_i|^{\lambda-1}\right], \quad \lambda\geq 1 \qquad (5)$$

or

$$\Delta x_i = \Delta t\,(x_{i-1} + x_{i+1} - 2x_i)|x_{i-1} + x_{i+1} - 2x_i|^{\lambda-1}\quad,\quad \lambda \geq 1. \qquad (6)$$

These equations are the unidimensional and discrete version of the following diffusion-like equations:[2]

$$\nabla \cdot \left(|\nabla x|^{\lambda-1}\nabla x\right) - \frac{\partial x}{\partial t} = 0 \qquad (7)$$

and

$$|\nabla^2 x|^{\lambda-1}\nabla^2 x - \frac{\partial x}{\partial t} = 0 \qquad (8)$$

that fall within the cases described by Eq. (1), which are described by the power law of Eq. (3) with $\gamma = \dfrac{1}{\lambda-1}$, $0 \leq \gamma \leq \infty$ (Ref. 11).

We have registered the time evolution of the values $x_n = X(n,t)$, $n = 0\ldots N-1$, starting from random initial configurations and we have calculated $\psi$ as the following average

$$\psi = \sqrt{\langle x^2 \rangle} = \sqrt{\frac{1}{N}\sum_{i=0}^{N-1} x_i^2}\ . \qquad (9)$$

To our surprise, *it is found that, as time runs, the configurations of the spatial values of the field X tend to an invariant shape $X_0$* (Figs. 1(a), (b) and (c)). The inspection of $\psi(t)$ shows that the stationary regime begins when a configuration with the shape of $X_0$ is attained (see Fig. 1(d)), indicating a clear link between both situations. After this we observe that the evolution of the system with time consists only of variations of scale of such curve. It seems therefore reasonable to simplify this situation by writing

$$X = \psi(t)\,X_0 \qquad (10)$$

i.e., identifying $\psi$ with the scalar magnitude measured in the process, as $\psi$ is an average of the values of $X$. Attending to this it is obvious that, once the stationary regime is attained, $X$ is an eigenfunction of $\Theta$ (i.e., $\Theta X = \psi'(t) X_0 = \psi'(t)/\psi(t)\,\psi(t) X_0 = \mu X$, $\mu$ a time-dependent scalar). The kind of time evolution described by Eq. (10) suggests that in the stationary state the evolution of the system tends to be a change of scale of the configurations of the field and it also suggests that scale effects could be a key component of the nature of the interaction described by the operator $\Theta$. This means at least that, if $X$ is an eigenfunction of $\Theta$, a scaling of $X$ must give rise to a scaling of $\Theta X$ even if the scaling relation is not the same. *It must be remarked that this does not mean that $\Theta$ is linear.*

Making use of this clue given by the simulations we have tried to see whether it is the consequence of a more basic property. On the basis of the aforementioned observations, a general, simple and sufficient condition for the existence of a function giving rise to the evolution described by Eq. (10) is that, for any scalar $\alpha$, there exists another scalar $\beta$ so that

$$\Theta \alpha = \beta \Theta \qquad (11)$$

(note that this condition applies only to the properties of the operator, not of the field **X**).

The central point of this letter is to see whether we can find a connection of this general condition with a stationary regime. For this purpose we have first to demonstrate the following lemma:

*Lemma*: Given an eigenfunction $\mathbf{X}_0$ of an operator $\Theta$ of the kind described by Eq. (10), there exists a real function $\psi(t)$ that fulfils the following equation:

$$\Theta(\psi \mathbf{X}_0) - \frac{\partial(\psi \mathbf{X}_0)}{\partial t} = 0 \qquad (12)$$

*Proof*: From Eq. (12) and taking into account Eq. (11) we have (note that $\mathbf{X}_0$ depends only on the spatial coordinates and it is not affected by the time derivative)

$$\psi'\mathbf{X}_0 = \frac{\partial(\psi\mathbf{X}_0)}{\partial t} = \Theta(\psi\mathbf{X}_0) = \beta\Theta\mathbf{X}_0 = \beta\mu\mathbf{X}_0 \qquad (13)$$

$\beta$ depends, obviously, on $\psi$, and $\mu$ is a constant. Therefore we arrive to the equation

$$\psi' = \mu\beta(\psi) \qquad (14)$$

which we suppose it has a solution. Therefore Eq. (12) is correct.

We use now this result for the proof of the following theorem.

*Theorem*: If $\Theta$ is an operator fulfilling Eq. (11), then the function $\psi(t)$ described in the previous lemma fulfils the property described in Eq. (2), which identifies a stationary process.

*Proof*: Let be $\psi_1 = \psi(t_1)$ and $\psi_2 = \psi(t_2)$, and another two values $\psi_3$ and $\psi_4$ with similar definition and with the condition

$$\frac{\psi_1}{\psi_2} = \frac{\psi_3}{\psi_4} \qquad (15)$$

This means that $\psi_3 = \alpha\psi_1$ and $\psi_4 = \alpha\psi_2$ (i.e., the same $\alpha$ in both cases).

Now, as $\Theta$ fulfils Eq. (11)

$$\Theta(\psi_3\mathbf{F}) = \Theta(\alpha\psi_1\mathbf{F}) = \beta\Theta(\psi_1\mathbf{F})$$
$$\Theta(\psi_4\mathbf{F}) = \Theta(\alpha\psi_2\mathbf{F}) = \beta\Theta(\psi_2\mathbf{F}) \qquad (16)$$

and taking into account the previous lemma we obtain that $\psi'_3 = \beta\psi'_1$ and $\psi'_4 = \beta\psi'_2$, with the same $\beta$ in both cases; therefore

$$\frac{\psi'_1}{\psi'_2} = \frac{\psi'_3}{\psi'_4} \qquad (17)$$

i.e.: there is an unique relationship between quotients of values of $\psi$ at times $t_1$ and $t_2$ ($\psi_1/\psi_2$) and quotients of the time derivatives of $\psi$ at the same times ($\dot{\psi}_1/\dot{\psi}_2$) independently of the times chosen ($t_1$ and $t_2$ or $t_3$ and $t_4$); in other words, for $\psi$ there exists a function $f$ that fulfils Eq. (2).

According to our previous results,[11] this means that in any first-order relaxing process, if the interaction governing the system (described by a spatial-like operator $\Theta$) fulfils the scale property described in Eq. (11), then the scalar magnitude $\psi$ relaxes following a stationary process described by a power law of the type described in Eq. (3). This is coherent with the results found in some specific research areas, like for example the description of the temporal occurrence of earthquakes. Recently, the description of the time evolution of aftershocks given by Omori's law[9] has been completed after the discovery of scaling laws that involve the waiting times between earthquakes[12] and the rate of seismic occurrence.[13]

The scaling property of $\Theta$ described by Eq. (11) has also some interesting implications because in such a case:

$$\frac{\partial(\alpha \mathbf{X})}{\partial t} = \Theta(\alpha \mathbf{X}) = \beta \Theta \mathbf{X} = \beta \frac{\partial \mathbf{X}}{\partial t} \quad \Rightarrow \quad \frac{\partial(\alpha \mathbf{X})}{\partial(\beta t)} = \frac{\partial \mathbf{X}}{\partial t}, \tag{18}$$

integrating from $t$ to $\infty$:

$$\alpha \mathbf{X}\big|_{\beta t} = \mathbf{X}\big|_t \tag{19}$$

and now, if $\psi$ is obtained from $\mathbf{X}$ by any standard operation like that of Eq. (9) then we have

$$\psi(\beta t) = \psi \circ \mathbf{X}\big|_{\beta t} = \psi \circ \left(\frac{1}{\alpha} \mathbf{X}\bigg|_t\right) = \frac{1}{\xi} \psi \circ \mathbf{X}\big|_t = \frac{1}{\xi} \psi(t) \tag{20}$$

(in the case of an average obtained by Eq. (9) then $\alpha=\xi$). Therefore this leads us to the following important property:

$$\xi\psi(\beta t) = \psi(t) \qquad (21)$$

with $\xi$ and $\beta$ constants. This means that a scaling in time involves a scaling in $\psi$.

We have tried to obtain a direct confirmation of this from our simulations. First we see that the operators $\Theta$ of the models of Eqs. (5) and (6) fulfil the property described by Eq. (11):

$$\Theta\alpha = \alpha^\lambda \Theta \qquad (22)$$

therefore, if our prediction is correct, their functions $\psi$ should fulfil the scale property described in Eq. (21). We have checked that this is the case, by comparing couples of relaxation curves: one of them starting from initial conditions an order of magnitude higher than the other (Fig. 2). It is observed that the rescaling in the values of $\psi$ leads to an equivalent rescaling in time.

We have also looked for characteristic invariants of stationary processes, as they are described by adimensional time–independent variables. In order to find it we first see what happens after changing the time origin in Eq. (3)

$$\psi_0\left(1+\frac{t}{\gamma\tau}\right)^{-\gamma} = \psi_0'\left(1+\frac{t-t_0}{\gamma'\tau'}\right)^{-\gamma'} \qquad (23)$$

We obtain that $\gamma = \gamma'$ just by examining this equation for high $t$; but neither $\psi_0$ nor $\tau$ are invariants after the change. For this reason they cannot be considered as characteristic parameters of the process. But now, if we work out a little bit the previous expression

$$\psi_0'\left(1+\frac{t-t_0}{\gamma\tau'}\right)^{-\gamma} = \psi_0'\left(1-\frac{t_0}{\gamma\tau'}\right)^{-\gamma}\left(1+\frac{t}{\gamma\tau'(1-t_0/\gamma\tau)}\right)^{-\gamma} \qquad (24)$$

we have

$$\psi_0 = \psi_0'\left(1-\frac{t_0}{\gamma\tau'}\right)^{-\gamma}$$
$$\tau = \tau'\left(1-\frac{t_0}{\gamma\tau'}\right) \qquad (25)$$

where

$$\Gamma = \psi_0^{1/\gamma}\,\tau \qquad (26)$$

is invariant. For our simulations we observe that this holds perfectly (Fig. 3, inset). Also any function

$$f(\gamma, \Gamma) \qquad (27)$$

will be invariant, for obvious reasons.

Concerning changes of the initial conditions, we observe also that neither $\psi_0$ nor $\tau$ are invariants. But making use of the definition of $\Gamma$ we can rewrite the process described in Eq. (3)

$$\psi = \left[\psi_0^{-\frac{1}{\gamma}}\left(1+\frac{t}{\gamma\tau}\right)\right]^{-\gamma} = \left(\psi_0^{-\frac{1}{\gamma}}+\frac{t}{\gamma\Gamma}\right)^{-\gamma} \qquad (28)$$

and we observe that, given that Eq. (3) was obtained with the hypothesis of independence from the initial conditions,[11] the asymptotic limit should not depend on the initial configuration $X|_{t=0}$. In Fig. 3 we observe how the stationary limit is indeed the same after these changes and $\Gamma$ is also invariant under them. In the special case of the Debye process ($\gamma = \infty$) we have that $\Gamma = \tau$.

We have also found that $\Gamma$ depends only on $\gamma$ and on the system length $N$ of the initial random sequence:

$$\Gamma = \Gamma_0 N^{\alpha} \qquad (29)$$

as seen in Fig. 4 (it does not hold for small $N$, although). The exponent $\alpha$ depends on $\gamma$ and it is different in each model. It is worth mentioning that for both models there seems to be an intersection point $(2\pi, 1)$, $\Gamma_0 = 1/2\pi$, which corresponds to the value $\Gamma(2\pi) = 1$ obtained for the linear continuous case, whose solution can be obtained analytically

$$X_0 = \cos(2\pi x/N + \phi), \qquad (30)$$

$\phi$ an arbitrary phase.

The independence on the initial conditions and on the time origin suggests that the stationary process is an intrinsic property of the system and that the initial conditions play an effect analogous to a shift of the time origin $t_0$; that is, they are defining the degree of evolution of the process. From this point of view, the value $\gamma\tau$ would be a sort of *elapsed time* since an effective origin of the process.

In summary, we have found that in systems that show first-order relaxation, if the interaction that describes the system follows the scale property defined by Eq. (11), then the relaxation will attain a stationary regime described by a power law. This result allows to predict the relaxation of many complex systems and also, in the inverse way, whenever a power law behavior is found experimentally, it suggests that it could be interesting to investigate whether the interactions that describe the system fulfil any scale property.

**References**

1. *Relaxation in complex systems*, K. L. Ngai and G. B. Wright, editors, (Office of naval research, Arlington, 1984).

2. A. Aharoni, *Introduction to the theory of ferromagnetism*, (Clarendon Press, Oxford, 1996), ch. 5.

3. K. Binder and A. P. Young, Rev. Mod. Phys. **58**, 801 (1986).

4. J. L. Dormann, D. Fiorani, and E. Tronc, Adv. Chem. Phys. **98**, 283 (1997).

5. J. García-Otero, M. Porto, J. Rivas, and A. Bunde, Phys. Rev. Lett. **84**, 167 (2000).

6. M. Ulrich, J. García-Otero, J. Rivas, and A. Bunde, Phys. Rev. B **67**, 024416 (2003).

7. X. Chen, W. Kleemann, O. Petracic, O. Sichelschmidt, S. Cardoso, and P. P. Freitas, Phys. Rev. B **68**, 54433 (2003).

8. X. Chen, S. Sahoo, W. Kleemann, S. Cardoso, and P. P. Freitas, Phys. Rev. B **70**, 172411 (2004).

9. F. Omori, J. Coll. Sci. Imper. Univ. Tokio **7**, 111 (1895).

10. D. Brockmann, L. Hufnagel, and T. Geisel, Nature **439**, 462 (2006).

11. A. Fondado, J. Mira, and J. Rivas, Phys. Rev. B **72**, 024302 (2005).

12. P. Bak, K. Christensen, L. Danon, and T. Scanlon, Phys. Rev. Lett. **88**, 178501 (2002).

13. A. Corral, Phys. Rev. Lett. **92**, 108501 (2004).

**Figure captions**

Figure 1: (Color online) (a), (b) and (c): Configurations of the spatial values of the simulations of the model described by Eq. (5) with $N= 64$ and $\lambda= 2$ at selected times, corresponding to several stages of the relaxation process. (d) Main frame: Plot $\psi$ of vs. time, marking the different time zones corresponding to the groups of configurations labelled as A, B and C. Inset: fit of the relaxation process to a power law (Eq. (3)) in the way described in Ref. 11 ($d\psi / (d \ln t)$ vs. $\psi$).

Figure 2: (Color online) Verification of the scale property defined by Eq. (21) in the model corresponding to Eq. (5) with $N= 64$ for two different values of $\lambda$. The random initial conditions $X|_{t=0}$ of one of the simulations (absolute values $|X|_{t=0}|$ between 0 and 1) are multiplied by a factor of 10 in the other (absolute values $|X|_{t=0}|$ between 0 and 10). The result is the predicted shift of the time scale of the curves.

Figure 3: (Color online) Main frame: Simulations with different random initial conditions of the processes of Fig. 2. Inset: $\tau$ vs $\psi_0$ of the processes described in Fig. 2 and of other (including those of the main frame) with different random initial conditions $X|_{t=0}$ and different amplitude scales. The invariance of the parameter $\Gamma= \psi_0^{1/\gamma} \tau$ is observed in all cases. Absolute values

Figure 4: (Color online) Dependence of $\Gamma$ on $N$ for several values $\lambda$ in the models described by (a) Eq. (5) and (b) by Eq. (6).

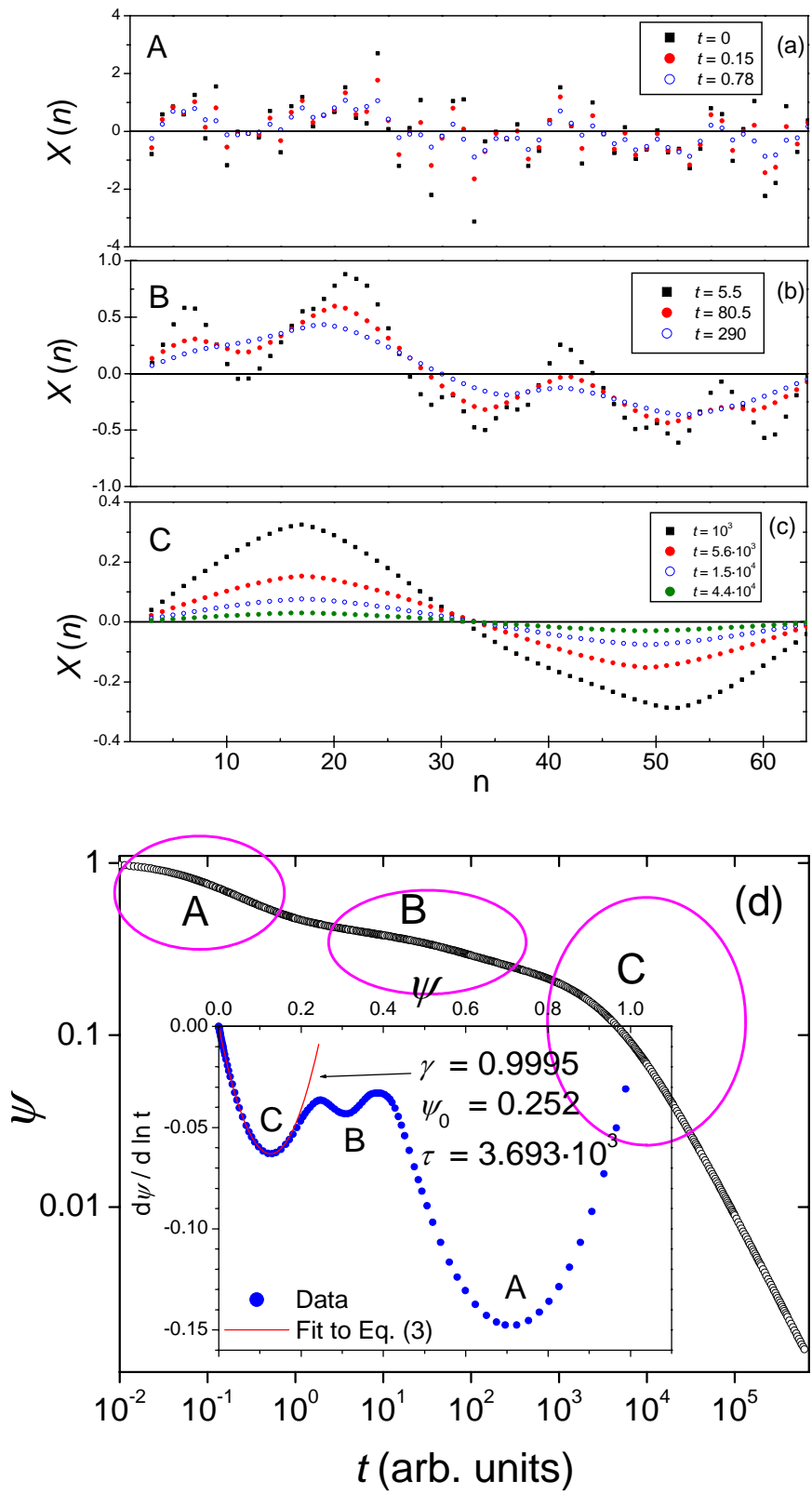

Figure 1

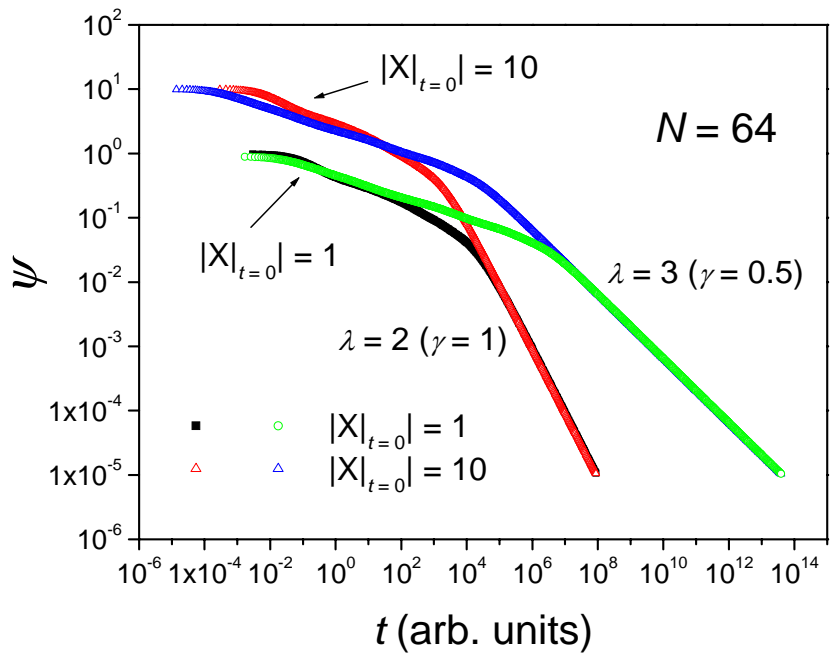

Figure 2

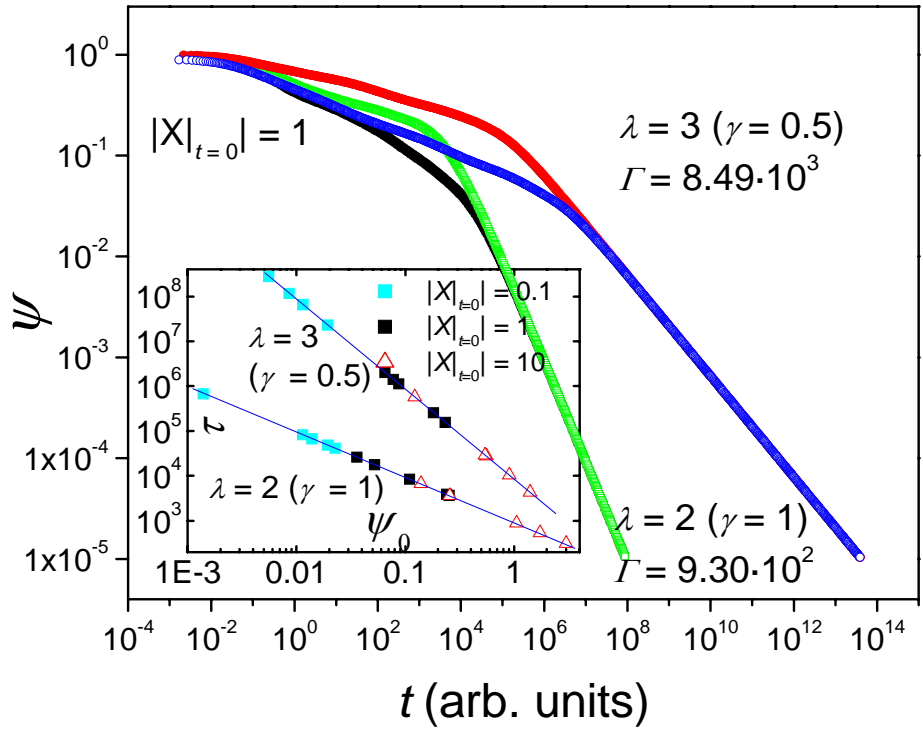

Figure 3

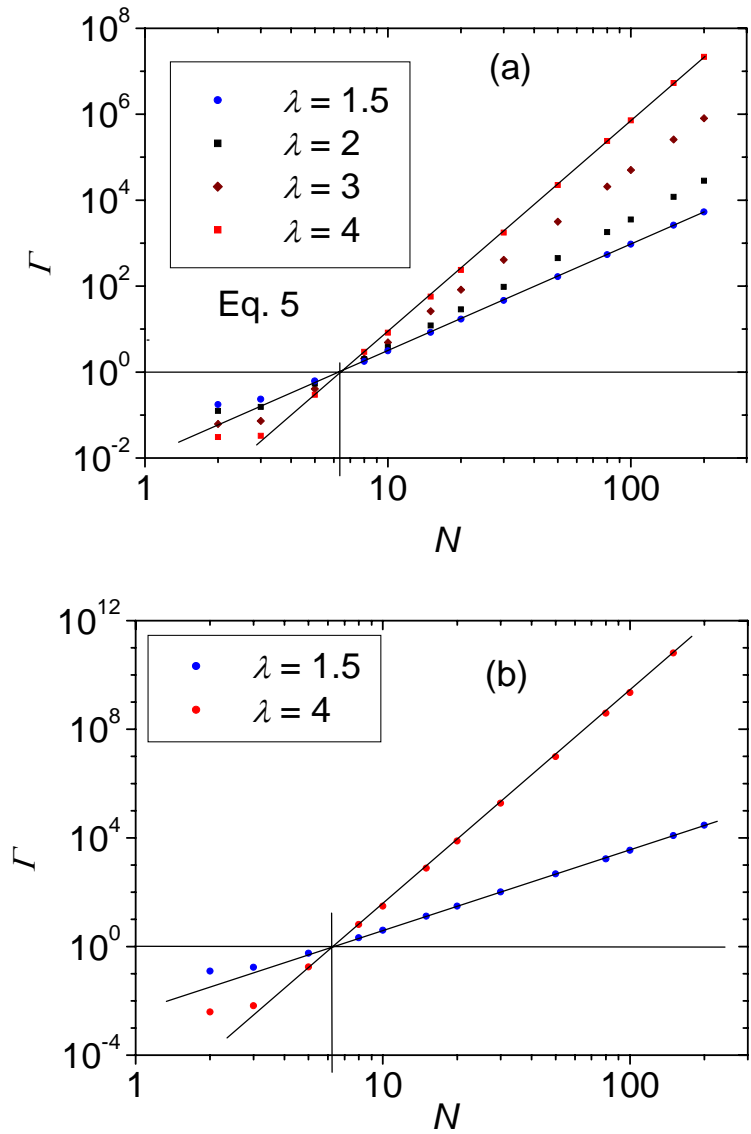

Figure 4